\documentclass{article}

\usepackage{microtype}
\usepackage{graphicx}
\usepackage{booktabs} % for professional tables

\usepackage{hyperref}

\usepackage[accepted]{mlsys2025}

\hypersetup{
    pdfsubject={Preprint}
}

\usepackage[subrefformat=parens]{subcaption}
\usepackage{enumitem}
\usepackage{xurl}
\usepackage[fleqn]{amsmath}

\mlsystitlerunning{Norn: Context-Length-Aware Optimization of Weight and KV Cache Placement for Offloaded LLM Inference}

\begin{document}

\twocolumn[
\mlsystitle{Norn: Context-Length-Aware Optimization of Weight and KV Cache Placement for Offloaded LLM Inference}

% unless you've provided the [accepted] option to the mlsys2025


\mlsyssetsymbol{equal}{*}

\begin{mlsysauthorlist}
\mlsysauthor{Naoto Sugiura}{keio}
% \mlsysauthor{Bauiu C.~Yyyy}{equal,to,goo}
\mlsysauthor{Hiroki Matsutani}{keio}
% \mlsysauthor{Iaesut Saoeu}{ed}
% \mlsysauthor{Fiuea Rrrr}{to}
% \mlsysauthor{Tateu H.~Yasehe}{ed,to,goo}
% \mlsysauthor{Aaoeu Iasoh}{goo}
% \mlsysauthor{Buiui Eueu}{ed}
% \mlsysauthor{Aeuia Zzzz}{ed}
% \mlsysauthor{Bieea C.~Yyyy}{to,goo}
% \mlsysauthor{Teoau Xxxx}{ed}
% \mlsysauthor{Eee Pppp}{ed}
\end{mlsysauthorlist}

\mlsysaffiliation{keio}{Keio University, Kanagawa, Japan}
% \mlsysaffiliation{goo}{Googol ShallowMind, New London, Michigan, USA}
% \mlsysaffiliation{ed}{School of Computation, University of Edenborrow, Edenborrow, United Kingdom}

\mlsyscorrespondingauthor{Naoto Sugiura}{ko1234@keio.jp}
% \mlsyscorrespondingauthor{Eee Pppp}{ep@eden.co.uk}

\mlsyskeywords{Large language model, Offloaded Inference, Scheduling}

\vskip 0.3in

\begin{abstract}
% This document provides a basic paper template and submission guidelines.
% Abstracts must be a single paragraph, ideally between 4--6 sentences long.
% Gross violations will trigger corrections at the camera-ready phase.

Offloading model weights and the key-value (KV) cache to host memory enables large language models (LLMs) to run on a single memory-constrained GPU. 
As the KV cache grows during decoding, GPU memory requirements and the costs of KV cache transfers and attention computation change. 
However, existing offloading methods generally use static computation assignments and placements of model weights and the KV cache, preventing them from adapting to these changes and degrading inference performance. 
We propose Norn, a context-length-aware offloaded LLM inference method. 
Based on offline profiling, Norn co-optimizes computation assignments and placements at the sublayer level, precomputing multiple execution plans and their switching points.
During decoding, Norn switches plans as context length increases, overlapping computation and data transfers without knowing the output length in advance. 
For prefill, Norn employs operator-major chunked prefill to avoid repeated weight transfers and automatically selects the chunk size. 
Experiments with Qwen3-14B and Qwen3-32B on two single-GPU systems show up to a 16.8$\times$ speedup in end-to-end latency over Pipelined Sharding and a 4.2$\times$ speedup over its oracle variant with known input and output lengths.

\end{abstract}
]



\printAffiliationsAndNotice{}  % leave blank if no need to mention equal contribution
% \printAffiliationsAndNotice{\mlsysEqualContribution} % otherwise use the standard text.

%%% equation 環境の前後の余白
\setlength{\abovedisplayskip}{3pt}
\setlength{\belowdisplayskip}{3pt}
\setlength{\abovedisplayshortskip}{3pt}
\setlength{\belowdisplayshortskip}{3pt}

\section{Introduction}

The use of LLMs has expanded beyond short conversations to complex reasoning for intricate problems and long-context processing of massive documents. 
As LLM applications continue to diversify, interest in running LLMs on local devices such as PCs and smartphones, rather than relying exclusively on cloud services, has grown. 
Security concerns regarding the external transmission of personal or confidential information are further fueling the demand for local inference~\cite{bloomberg}.
On-device language models have already enabled practical AI features such as writing assistance and summarization~\cite{li2025apple,pradeep2024phisilica}, while their use in latency-sensitive applications, including in-game AI, is also being explored~\cite{ukarande2026efficient}. 
In contrast to data-center environments that primarily prioritize high throughput, local environments emphasize processing small batch sizes with low latency.

However, running LLMs with large parameter counts requires substantial GPU memory, making the limited memory capacity of consumer-grade GPUs a major challenge. 
During inference, GPU memory is consumed not only by model weights but also by the KV cache, which retains past token information.
While the memory footprint of weights remains constant, that of the KV cache grows linearly with context length. 
For example, with FP16 precision, the KV cache of Qwen3-32B~\cite{yang2025qwen3} alone reaches approximately 32~GiB at 128K tokens, comparable to the memory capacity of a consumer-grade GPU. 
% KV cache requirements vary substantially across short conversations, reasoning tasks that generate long outputs, and long-context tasks that process large inputs. 
Furthermore, KV cache requirements vary substantially with the input and output lengths of each request.
In particular, during long-output generation, the final KV cache size is uncertain because the output length is generally unknown in advance.
% Unlike the input length, the output length is generally unknown at the start of inference, making the final KV cache size uncertain.
% This makes it difficult to determine an appropriate GPU memory allocation at the start of inference. 
% Handling a mix of requests with different memory requirements requires configuring memory capacity per request. 
% As users submit requests with different memory requirements, they must configure GPU memory allocation for each request.
% Therefore, local LLM inference must support requests with diverse input and output lengths under limited GPU memory capacity while maintaining low inference latency.

A promising approach to mitigating GPU memory constraints is offloaded inference, which leverages host memory. 
In offloaded inference, a portion of the model weights and KV cache is stored in host memory and transferred to the GPU as needed, enabling the execution of LLMs that exceed the available GPU memory capacity. 
However, data transfers between the CPU and GPU introduce additional overhead, making offloading alone insufficient to achieve high inference performance. 
To address this challenge, several methods have been proposed to reduce inference latency by optimizing the assignment of computation to the CPU or GPU and the placement of weights and the KV cache, while overlapping computation with data transfers~\cite{yu2024twinpilots,ukarande2026efficient,sugiura2026layer,park2024improving,zhao2024hetegen}.

However, existing offloading methods do not consider adapting computation assignments or placement as the context length increases during decoding. 
% ~\cite{sugiura2026layer,ukarande2026efficient}. 
For example, Pipelined Sharding~\cite{ukarande2026efficient} selects an execution plan based on the number of new tokens to process, but does not adapt placement in response to context-length growth during decoding with a batch size of one.
Static placement introduces a trade-off between the maximum supported context length and inference performance. Reserving GPU memory for the KV cache to accommodate long contexts reduces memory available for resident weights, degrading inference performance even for short requests. 
Conversely, a placement tailored to short contexts allows more weights and the KV cache to reside on the GPU, but the growing KV cache may exhaust GPU memory during generation, preventing inference from continuing under the same placement.
In our measurements using Pipelined Sharding with Qwen3-14B, increasing the maximum supported context length from 2K to 128K tokens increased per-token decoding latency by 18.7$\times$, even with fixed input and output lengths (Figure~\ref{fig:back-ps}).
To avoid excessive GPU memory reservation for short requests, users would need to configure the maximum supported context length for each request, even though the output length is unknown in advance.

Furthermore, increasing context length affects not only memory requirements but also the efficiency of offloaded inference. 
While KV cache transfer time and attention computation time increase with context length, weight transfer time and Feed-Forward Network (FFN) computation time remain 
% approximately 
constant. 
Consequently, the relative costs of computation and data transfer change, potentially altering the optimal computation assignments and placement during inference.
Thus, a placement that is efficient at one context length may become suboptimal as decoding progresses. 
% To achieve low-latency inference across requests with diverse input and output lengths, the placement of weights and the KV cache, as well as the assignment of computation to the CPU or GPU, must adapt to changes in context length.

To address these challenges, we propose Norn, an offloaded inference method that dynamically switches execution plans as context length increases. 
Using offline profiling results, Norn co-optimizes the assignment of attention and FFN sublayers to CPU or GPU execution and the placement of weights and the KV cache. 
% It determines multiple execution plans and their switching points in advance, considering context-length-dependent performance changes and the data migration costs associated with plan switching. 
It determines multiple execution plans and their switching points in advance, accounting for context-dependent performance changes and data migration costs. 
During decoding, Norn switches plans, adjusts GPU memory allocation, and overlaps computation with data transfers.
% During inference, Norn switches between these plans as the context length increases and allocates GPU memory for the KV cache according to the context-length range covered by each plan. 
% Norn also asynchronously transfers weights and the KV cache that are not resident on the GPU, overlapping computation with data transfers. 
% This approach avoids excessive GPU memory reservation for future KV cache growth while adapting to changes in computation and transfer costs as the context length increases, thereby achieving low-latency inference.
For prefill, Norn employs operator-major chunked prefill to reduce repeated weight transfers while limiting activation memory usage, and automatically selects the chunk size considering prefill latency and the cost of transitioning to decoding.
We evaluated Norn using Qwen3-14B and Qwen3-32B on two desktop systems. Norn achieved up to a 16.8$\times$ speedup in end-to-end latency over Pipelined Sharding and up to a 4.2$\times$ speedup over its oracle configuration in which input and output lengths are known in advance.

In summary, this work offers the following contributions:
\begin{itemize}[nosep, leftmargin=1.2em, labelsep=0.4em]
\vspace{-2mm}
    \item We show that static placement degrades short-context decoding performance by reserving GPU memory for future KV cache growth.
    \item We propose a dynamic planning method that precomputes multiple execution plans with optimized computation assignments and placements, along with their switching points.
    \item We propose operator-major chunked prefill with automatic chunk-size selection to reduce repeated weight transfers while limiting memory usage and prefill-to-decoding transition overhead.
    \item We evaluate Norn on two GPU systems with Qwen3-14B and Qwen3-32B, demonstrating lower inference latency than Pipelined Sharding across diverse settings.
\end{itemize}

\section{Background and Motivation}

\subsection{Scheduling for Offloaded LLM Inference}

In offloaded LLM inference, model weights and the KV cache can be placed in either GPU memory or host memory, while computation can be assigned to the CPU or GPU. 
GPU execution can use either resident execution, which keeps weights in GPU memory, or streaming execution, which stores weights in host memory and transfers them to the GPU as needed. 
Similarly, the KV cache accessed by attention can either reside in GPU memory or be transferred from host memory during execution.

Transferring weights and the KV cache required for subsequent GPU computation while other computation is in progress reduces inference latency. 
Existing methods optimize the assignment of computation to the CPU or GPU and the placement of weights and the KV cache while accounting for this overlap. %~\cite{yu2024twinpilots,ukarande2026efficient,sugiura2026layer}. 
For example, Layer-wise Scheduling~\cite{sugiura2026layer} assigns each layer to CPU execution, GPU resident execution, or GPU streaming execution and maximizes overlap by transferring the weights of subsequent layers during CPU computation. 
Pipelined Sharding~\cite{ukarande2026efficient} determines computation assignments and placement at the granularity of attention and FFN sublayers, prioritizing GPU residency for attention weights and the KV cache. 
In execution plans that combine CPU computation with GPU streaming, it prefetches the weights of the next sublayer during CPU computation to overlap computation with data transfers.

\subsection{Changes in Computation and Transfer Costs}
\label{ssec:perf}

\begin{figure}[t]
    \centering
    \begin{minipage}{0.49\columnwidth}
        % \vspace{-4mm}
        \includegraphics[width=\linewidth]{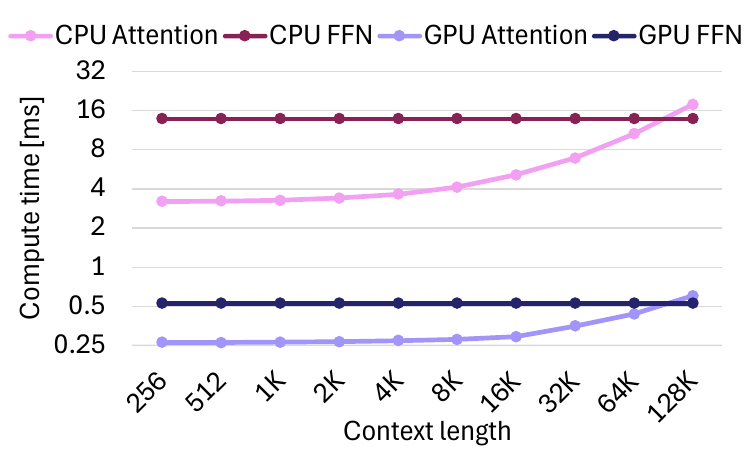}
        \vspace{-5mm}
        \subcaption{Computation time.}
        % \label{fig:ex1-boxplot}
    \end{minipage}
    \hfill % 画像間のスペース
    \begin{minipage}{0.49\columnwidth}
        \includegraphics[width=\linewidth]{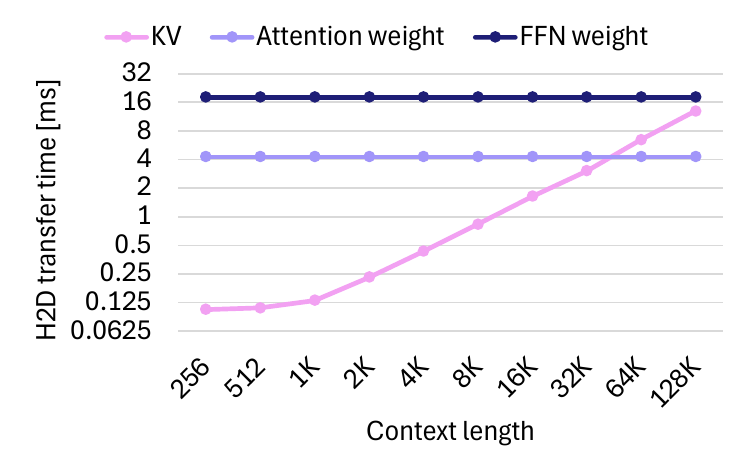}
        \vspace{-5mm}
        \subcaption{Transfer time.}
        % \label{fig:ex2-update}
    \end{minipage}
    \hfill % 画像間のスペース
    \vspace{-2mm}
    \caption{Computation time and transfer time per sublayer of Qwen3-32B.}
    \label{fig:back-perf}
    \vspace{-4mm}
\end{figure}

Changes in computation and transfer costs with increasing context length are not uniform. 
Figure~\ref{fig:back-perf} shows the computation and transfer times per sublayer for Qwen3-32B. 
As the context length increases from 256 to 128K tokens, attention computation time increases by 5.6$\times$ on the CPU and 2.3$\times$ on the GPU. 
Since attention accesses the entire KV cache of preceding tokens, its computational cost increases with context length. 
Furthermore, the amount of KV cache transferred grows linearly with context length, resulting in a 123$\times$ increase in transfer time. 
In contrast, FFN computation times on both the CPU and GPU remain constant, and sublayer weight transfer times are independent of context length. 
Similar trends were observed for Qwen3-14B. 
These changes in the relative costs of computation and data transfers suggest that efficient computation assignments and placement may vary with context length.

\subsection{Limitations of Static Placement}

\begin{figure}[t]
    \centering
    \begin{minipage}{0.49\columnwidth}
        % \vspace{-4mm}
        \includegraphics[width=\linewidth]{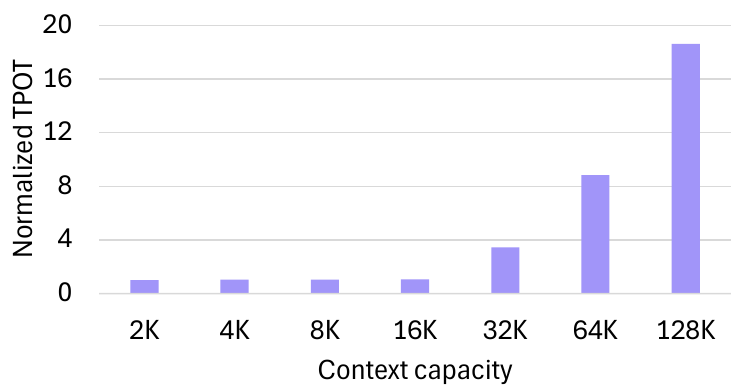}
        \vspace{-5mm}
        \subcaption{Qwen3-14B.}
        % \label{fig:ex1-boxplot}
    \end{minipage}
    \hfill % 画像間のスペース
    \begin{minipage}{0.49\columnwidth}
        \includegraphics[width=\linewidth]{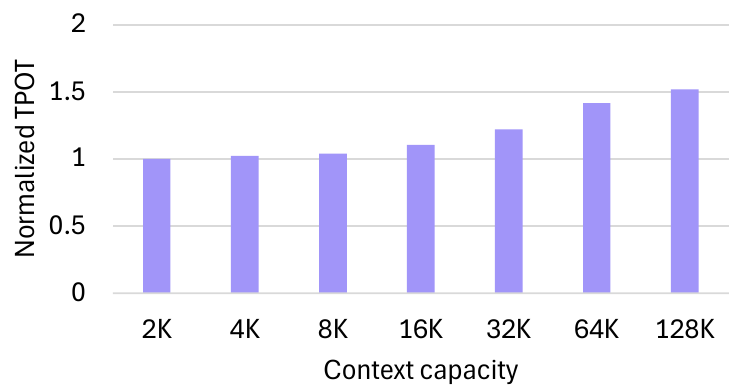}
        \vspace{-5mm}
        \subcaption{Qwen3-32B.}
        % \label{fig:ex2-update}
    \end{minipage}
    \hfill % 画像間のスペース
    \vspace{-2mm}
    \caption{Impact of fixed context capacity on TPOT. TPOT is normalized to the 2K configuration for each model.}
    \label{fig:back-ps}
    \vspace{-6mm}
\end{figure}

Most existing methods do not change the assignment of computation to the CPU or GPU or the placement of weights and the KV cache during decoding. 
Since the KV cache grows linearly with context length, static placement requires GPU memory to be allocated in anticipation of future KV cache growth. 
For example, Pipelined Sharding and llama.cpp~\cite{llamacpp} require the maximum supported context length to be specified. 
Although Hugging Face Accelerate~\cite{accel} and Layer-wise Scheduling do not require specifying a maximum context length, the amount of weights placed on the GPU must still be limited to accommodate KV cache growth.

Such static placement causes performance problems both during the decoding of a single request and across requests with different context lengths. 
First, for requests that generate long outputs, such as reasoning tasks, decoding performance degrades in the early stages of generation. 
Static placement reserves GPU memory for the KV cache expected near the end of generation, limiting the amount of weights that can remain resident on the GPU even when the KV cache is small at the beginning. 
Consequently, CPU computation and weight transfers introduce performance overhead compared with a placement optimized for shorter contexts. 
Furthermore, as discussed in Section~\ref{ssec:perf}, changes in computation and transfer costs mean that a single placement may not remain efficient throughout decoding. 
% Second, performance also degrades when a workload contains requests with different context lengths. 
% Allocating GPU memory to support long requests increases weight offloading even for short requests, thereby degrading inference performance. 
Second, when the same static placement is used for requests with different context lengths, GPU memory must be reserved to accommodate long requests. 
This reduces the memory available for GPU-resident weights even for short requests, degrading inference performance.
Conversely, a placement tailored to short contexts may lead to GPU memory exhaustion as the KV cache grows in longer requests, preventing inference from continuing. 
One possible approach is to specify the maximum context length for each request. 
However, since the actual output length is unknown in advance, it is impractical to expect users to manually select an appropriate setting for each request.

To quantify this effect, we evaluated Pipelined Sharding with input and output lengths fixed at 1024 and 32 tokens, respectively, while varying only the maximum supported context length from 2K to 128K tokens. Figure~\ref{fig:back-ps} shows the results. For Qwen3-14B, time per output token (TPOT) changed little up to the 16K setting but increased to approximately 3.4$\times$ and 18.7$\times$ that of the 2K setting at 32K and 128K, respectively. This is because all weights and the KV cache can reside on the GPU with the 2K setting, whereas increasing the maximum supported context length requires more weights to be offloaded to reserve GPU memory for the KV cache. For Qwen3-32B, TPOT also increased by 51.8\% at the 128K setting compared with the 2K setting. These results demonstrate that static placement based on future KV cache requirements can degrade decoding performance even when the actual context length being processed is identical. 
In particular, even if an appropriate maximum context length could be specified for each request, using the same placement throughout generation would prevent adaptation to performance changes during decoding. Therefore, the placement of weights and the KV cache must be dynamically adjusted as context length increases.

\subsection{Challenges in Dynamic Placement}

Dynamic placement requires execution plan generation and the determination of switching points to be considered jointly. 
Each execution plan must accommodate the growing KV cache until the next switch, and the GPU memory reserved for this purpose constrains weight placement and inference performance. 
Therefore, the choice of placement depends on the context length up to which that placement will be used. 
In practice, requests have diverse input and output lengths, and the output length is unknown at the start of inference, making these decisions even more challenging.

Generating execution plans involves a combinatorial placement search problem. 
For each attention and FFN sublayer, the assignment of computation to the CPU or GPU and the placement of weights and the KV cache must be determined, and the number of candidate plans grows combinatorially with the number of sublayers. 
TwinPilots~\cite{yu2024twinpilots} formulates the assignment of operations to the CPU and GPU as an NP-complete load balancing problem. 
Repeating such a search during decoding could introduce substantial computational overhead.

Determining switching points requires considering not only GPU memory constraints but also changes in the relative performance of execution plans as context length increases. 
Even if an execution plan satisfies GPU memory constraints, another plan may become faster as context length increases. 
However, switching plans involves migrating weights and the KV cache, and frequent switching increases migration costs. 
Therefore, rather than independently selecting the fastest placement at each context length, the set of execution plans and their applicable context-length ranges must be determined jointly, considering the execution performance of each plan and the costs of switching between plans.

\section{Proposed method}
\label{sec:method}

% なぜオフラインで決めるか？

\subsection{Overview of Norn}

\begin{figure}[t]
  \centering
  \includegraphics[width=\columnwidth]{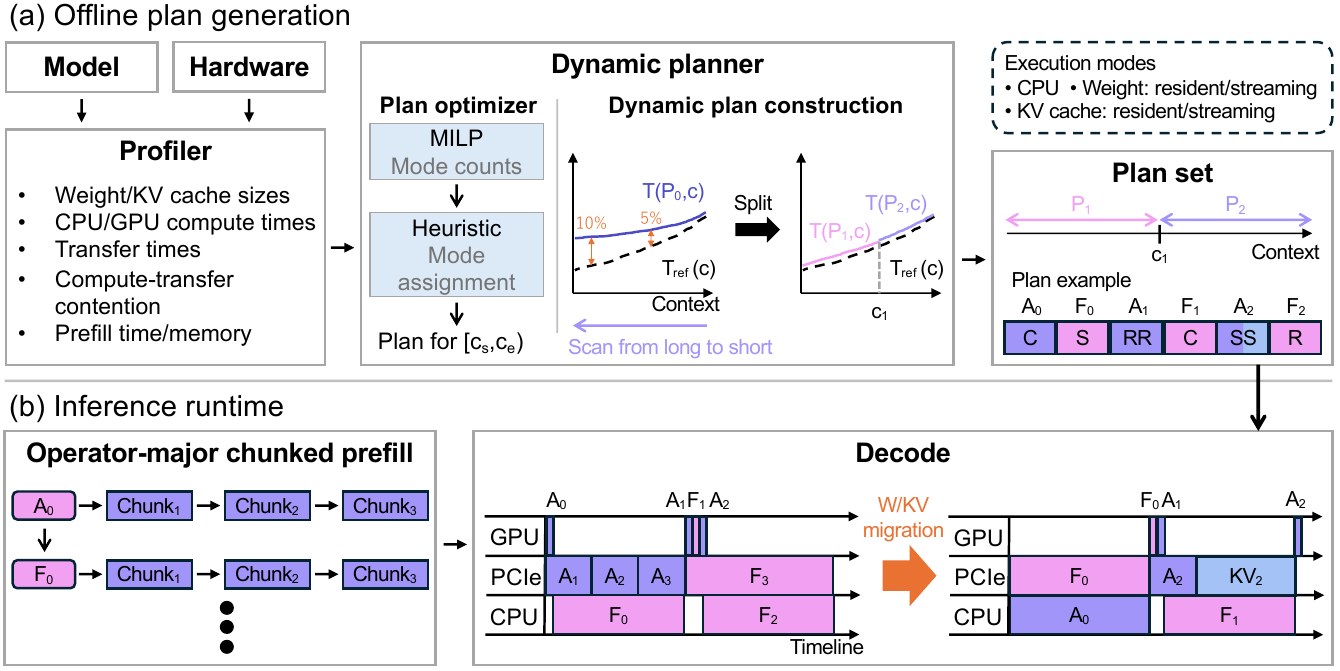}
  \vspace{-7mm}
  \caption{Overview of Norn.}
  \label{fig:overview}
  \vspace{-7mm}
\end{figure}

Norn is an offloaded inference method that dynamically switches execution plans as the context length increases.
As shown in Figure~\ref{fig:overview}, Norn consists of three components, the Profiler, Dynamic Planner, and Inference Runtime.
The Profiler measures computation times, data transfer times, and compute-transfer contention. %on the target hardware.
Based on the profiling results, the Dynamic Planner determines multiple execution plans and their switching points offline for different context lengths.
For prefill, the Inference Runtime performs GPU computation and applies operator-major chunked prefill to long inputs.
During decoding, it overlaps computation with data transfers according to each plan and migrates the necessary data when the context length reaches a switching point.
This enables dynamic placement without predicting output lengths or reoptimizing execution plans during inference.

\subsection{Profiler}

The Profiler measures computation and transfer times on the target hardware to optimize performance for a given model and execution environment.
It also obtains the weight size of each sublayer and the KV cache size per token from the model configuration.
For decoding, it measures CPU and GPU execution times for each sublayer, transfer times for weights, the KV cache, and activations, compute-transfer contention, and host-side overhead of launching GPU operations.
Attention computation times and KV cache transfer times, which depend on context length, are measured at multiple context lengths and linearly interpolated between measurement points.
For prefill, it measures GPU execution times for each sublayer and peak GPU memory usage across multiple input lengths and chunk sizes.

\subsection{Dynamic Planner}
\label{sec:dynamic_planner}

The Dynamic Planner consists of optimizing individual execution plans using the Plan Optimizer and determining a set of execution plans and their switching points while accounting for performance degradation and switching costs.

\subsubsection{Plan Optimizer}
\label{sssec:plan_optimizer}

The Plan Optimizer generates a single execution plan $P$ for a context-length range $I=[c_s,c_e)$ with the objective of minimizing the estimated total decoding time.
Plan $P$ consists of the execution modes of attention and FFN in each layer and the buffer capacity $B$ used for transferring weights and the KV cache.
For FFN, three execution modes are available, namely CPU execution, GPU resident execution, and GPU streaming execution.
For attention, four GPU execution modes are defined by independently selecting resident or streaming placement for weights and the KV cache, in addition to CPU execution.
% Thus, attention has five possible modes, while FFN has three.

% The Plan Optimizer separately optimizes the number of sublayers assigned to each mode and their assignment order across layers.
The Plan Optimizer separately optimizes mode counts and their assignment order across layers.
For a model with $L$ layers, there are up to $15^L$ combinations of execution modes.
Furthermore, the overlap between computation and data transfers depends on temporal constraints, including the order of mode assignments and the start and completion times of transfers, making an exact optimization formulation computationally expensive.
However, once the number of sublayers assigned to each mode is determined, GPU memory usage and total computation and transfer times can be estimated.
Therefore, the Plan Optimizer first determines the mode counts using mixed-integer linear programming (MILP) and then determines their assignment order through a simulator-based heuristic search.
% This approach reduces the number of integer variables in the MILP while accounting for execution-order-dependent overlap between computation and data transfers.
This reduces the number of MILP integer variables while accounting for execution-order-dependent compute-transfer overlap.

%%%%%%
% MILP定式化
%%%%%%

The MILP minimizes a lower bound on decoding time estimated from the profiling results.
Evaluating every decoding step in the context-length range individually would be computationally expensive.
Therefore, the range is divided into $S$ equally spaced intervals, and the midpoint $c_i$ of each interval is used as a representative point.
Each representative point is assigned a weight $w_i$ corresponding to the number of decoding steps it represents.
Let $\tau_i$ denote the lower bound on TPOT at $c_i$.
The objective function is:
\begin{equation}
\min \sum_{i=1}^{S}w_i\tau_i
\end{equation}
%%%%%%%%%% 本当は段落分けてる %%%%%%%%%%

For each sublayer type, the mode counts must sum to the number of layers $L$. 
Let $\mathcal{Q}=\{\mathrm{Attention},\mathrm{FFN}\}$ denote the set of sublayer types and $\mathcal{M}_q$ the set of available modes for type $q\in\mathcal{Q}$. The integer variable $x_{q,m}\in\{0,\ldots,L\}$ denotes the number of sublayers of type $q$ assigned to mode $m$. 
The resulting constraint is:
\begin{equation}
\sum_{m\in\mathcal{M}_q}x_{q,m}=L
\end{equation}
The GPU memory constraint accounts for KV cache growth during decoding by using the resident memory requirements of weights and the KV cache at the end of the context-length range.
Let $R_{q,m}(c_e)$ denote the GPU resident memory required by a sublayer of type $q$ in mode $m$ at context length $c_e$, $M_{\mathrm{base}}$ the fixed GPU memory usage, and $M_{\mathrm{GPU}}$ the GPU memory capacity.
The resulting constraint is:
\begin{equation}
M_{\mathrm{base}}+B+
\sum_{q\in\mathcal{Q}}\sum_{m\in\mathcal{M}_q}
x_{q,m}R_{q,m}(c_e)
\leq M_{\mathrm{GPU}}
\end{equation}
Modes whose transfer data exceed $B$ are excluded, and host memory capacity constraints are imposed when necessary.
% Modes whose data to be transferred cannot fit within buffer $B$ are excluded, and host memory capacity constraints are imposed when necessary.

At each representative point $c_i$, let $C_i(\mathbf{x})$, $H_i(\mathbf{x})$, and $D_i(\mathbf{x})$ denote the total standalone execution times for computation, host-to-device (H2D) transfers, and device-to-host (D2H) transfers, respectively, based on the mode counts $\mathbf{x}$.
% The following basic lower-bound constraints are imposed on TPOT:
The following lower bounds are imposed on TPOT:
\begin{equation}
\tau_i \ge C_i(\mathbf{x}),\quad\tau_i \ge H_i(\mathbf{x}),\quad\tau_i \ge D_i(\mathbf{x})
\end{equation}
%%%%%%%%%% 本当は段落分けてる %%%%%%%%%%

To account for compute-transfer contention, the effect of overlapping execution on execution time is approximated.
Let $C_{i,a}(\mathbf{x})$ and $H_{i,b}(\mathbf{x})$ denote the total computation time for computation type $a$ and the total transfer time for transfer type $b$, respectively.
These quantities aggregate the standalone execution times according to the mode counts.
% Let $z_{i,a,b}\geq0$ denote the overlap duration, and $r^C_{i,a,b}$ and $r^H_{i,a,b}$ denote the relative processing rates under contention estimated from the profiling results.
% These rates are normalized to 1 for standalone execution.
Let $z_{i,a,b}\geq0$ denote the overlap duration, and let $r^C_{i,a,b}$ and $r^H_{i,a,b}$ denote the relative processing rates of computation and H2D transfers, respectively, under contention. These rates are estimated from the profiling results and normalized so that each equals 1 for standalone execution.
% Using $\gamma_{i,a,b}=r^C_{i,a,b}+r^H_{i,a,b}-1$, which represents the reduction in sequential execution time per unit of overlapping execution, the following constraints are added:
Using $\gamma_{i,a,b}=r^C_{i,a,b}+r^H_{i,a,b}-1$, which represents the difference between sequential and overlapped execution times per unit of overlap duration, the following constraints are added:
\begin{align}
&\sum_b r^C_{i,a,b}z_{i,a,b}
\leq C_{i,a}(\mathbf{x})\\
&\sum_a r^H_{i,a,b}z_{i,a,b}
\leq H_{i,b}(\mathbf{x})\\
&\tau_i
\geq C_i(\mathbf{x})+H_i(\mathbf{x})
-\sum_{a,b}\gamma_{i,a,b}z_{i,a,b}
\end{align}
During the overlap duration $z_{i,a,b}$, the amounts of computation and transfer work completed are represented by $r^C_{i,a,b}z_{i,a,b}$ and $r^H_{i,a,b}z_{i,a,b}$, respectively.
% The first two constraints ensure that the total work completed during overlapping execution does not exceed the total work required for each type.
The first two constraints bound the work completed during overlap by the total work required for each type.
The final constraint establishes a lower bound on TPOT by accounting for the time saved through overlapping execution.

The MILP thus determines the mode counts without directly assigning modes to individual layers. 
Directly modeling the mode selection for each layer would require $8L$ binary variables.
In contrast, this formulation uses only eight integer variables representing the mode counts. 
Since the mode counts for each sublayer type sum to $L$, the number of possible combinations is $O(L^6)$, reducing the exponential search space of mode assignments.

%%%%%%
% ヒューリスティック
%%%%%%

With the mode counts fixed by the MILP, the Plan Optimizer determines their assignment order across layers through a simulator-based heuristic search to account for compute-transfer overlap.
For mode assignment, the total execution time of plan $P$ is approximated as $\widehat{T}(P,I)=\sum_{i=1}^{S}w_iT(P,c_i)$, which is used as the evaluation function.
Here, $T(P,c_i)$ denotes the TPOT estimated by the simulator, accounting for dependencies between sublayers and the execution schedule of computation and data transfers.

First, an initial assignment is generated separately for attention and FFN using smooth weighted round-robin, with the mode counts as weights, to distribute the modes across layers.
Specifically, the mode with the largest deficit in cumulative assignments relative to its target proportion is selected.
Next, the mode sequences for attention and FFN are cyclically shifted independently to adjust their relative assignment order.
With one mode sequence fixed, cyclic shifts of the other are evaluated using the simulator, and the assignment minimizing $\widehat{T}(P,I)$ is selected.
This procedure is repeated alternately for attention and FFN.

Next, a local search is performed by swapping the execution modes assigned to two different layers of the same sublayer type.
During the search, candidate assignments are compared using $\widehat{T}(P,I)$, and worsening swaps are accepted with a certain probability to avoid becoming trapped in local optima.
The best plan found during the search is retained and selected.
Since swapping modes does not change the mode counts, the GPU resident memory requirements determined by the MILP are preserved.
% Feasibility is also verified at both ends of the context-length range.

Multiple candidates are also explored for the transfer buffer capacity $B$.
Increasing $B$ provides more room for prefetching but reduces the GPU memory available for resident weights and the KV cache.
Therefore, candidate capacities are generated based on capacity thresholds derived from the sizes of transferred data and buffer shortages detected through simulation.
For each candidate, the MILP and assignment-order optimization are performed.
Finally, the feasible plan that minimizes $\widehat{T}(P,I)$ is selected.

\subsubsection{Dynamic Plan Construction}

The objective of the Dynamic Planner is to minimize the sum of the total decoding time over the context-length ranges assigned to each plan and the data migration time incurred by plan switching.
% The Dynamic Planner recursively partitions the context-length range, then adjusts the switching points to account for switching costs and merges unnecessary ranges.
The Dynamic Planner recursively partitions the context-length range, then adjusts the switching points to account for switching costs and merges adjacent ranges to eliminate unnecessary plan switches.
Range partitioning is based on the performance degradation of each plan relative to a reference latency curve.
As context length increases, both attention computation time and KV cache transfer time increase, so an increase in TPOT alone cannot indicate the potential benefit of switching plans.
Therefore, the minimum TPOT among feasible reference plans at each context length is used as a baseline, and regions with substantial performance degradation are considered candidates for partitioning.

First, the Dynamic Planner constructs a reference latency curve as the basis for evaluating plan performance degradation. 
It selects multiple reference points from the target context-length range and uses the Plan Optimizer to generate a point plan for each reference point.
A point plan minimizes TPOT at its corresponding reference point.
Let $\mathcal{P}_{\mathrm{ref}}$ denote the set of reference plans.
The reference latency at context length $c$ is defined as $T_{\mathrm{ref}}(c)=\min_{P\in\mathcal{P}_{\mathrm{ref}}}T(P,c)$.
% To keep the evaluation baseline fixed throughout partitioning, the reference curve is not updated. %when additional plans are generated.
The relative performance degradation of plan $P$ with respect to the reference curve is defined as $\delta(P,c) = \frac{T(P,c)-T_{\mathrm{ref}}(c)}{T_{\mathrm{ref}}(c)}$.

The Dynamic Planner partitions the context-length range from longer to shorter contexts.
This is because the size of the KV cache increases monotonically with context length.
If a plan satisfies the GPU memory constraints at the end of a range, it remains feasible at shorter context lengths.
In contrast, determining plans from shorter to longer contexts requires reserving GPU memory for the KV cache based on how long each plan will remain in use.
If the reserved capacity is insufficient, an unplanned plan switch becomes necessary.
Since the reserved capacity also affects plan performance, determining an appropriate capacity is difficult.
By determining ranges from longer to shorter contexts, the planner can allocate memory based on the known end of each range, thereby avoiding unplanned switches caused by KV cache growth.

The Dynamic Planner recursively partitions a plan's applicable range if its performance degradation relative to the reference curve exceeds 10\% within that range.
For range partitioning, it scans each plan's applicable range from longer to shorter context lengths and identifies the first point where the performance degradation relative to the reference curve exceeds 5\%.
It then continues scanning toward shorter contexts to identify the first point where the degradation exceeds 10\%.
The 10\% threshold is used to determine whether partitioning is necessary, while the 5\% threshold is used to determine the partitioning point.
%%%%
Based on this partitioning rule, the Dynamic Planner first generates an initial plan covering the entire target context-length range $I=[1,c_{\max})$.
To prevent $c_{\max}$ from becoming the starting point for searching for a partitioning point, the performance degradation at $c_{\max}$ is constrained to be within 5\%.
If this condition is not satisfied, the optimization range is changed from the entire range $I$ to the latter half $I'=[(1+c_{\max})/2,c_{\max})$, and the plan is regenerated.
Due to the monotonic increase in memory usage with context length, the regenerated plan remains feasible over the original range $I$.
Next, the partitioning rule is applied to the initial plan.
If partitioning is required, the range is divided at a provisional switching point, and the Plan Optimizer regenerates plans for the resulting ranges.
This process is repeated until no further performance degradation requiring partitioning remains.

The switching points obtained through recursive range partitioning are provisional boundaries based on performance degradation and do not necessarily minimize the total execution time.
Therefore, the Dynamic Planner compares the decoding times of adjacent plans, proceeding from shorter to longer context lengths, and adjusts the switching points to where their TPOT curves intersect or to the boundaries of their feasible ranges imposed by memory capacity.
When a boundary is adjusted, the plans are regenerated for their new applicable ranges, and the adjustment is accepted only if it reduces the total execution time, including the data migration time incurred by plan switching.

Furthermore, the reduction in decoding time achieved by range partitioning may not justify the data migration cost incurred by plan switching.
Therefore, the Dynamic Planner generates a new plan for the combined range of two adjacent ranges and compares the total execution time before and after merging.
This comparison accounts for both the decoding time in each range and the transfer time for weights and the KV cache incurred by plan switching.
If merging reduces the total execution time, the switching point between the two ranges is removed.
The final set of execution plans includes each plan's applicable context-length range, the execution mode of each sublayer, the transfer buffer capacity, and data migration information required for switching between plans.

\subsection{Inference Runtime}
\label{sec:inference_runtime}

\begin{figure}[t]
  \centering
  \includegraphics[width=\columnwidth]{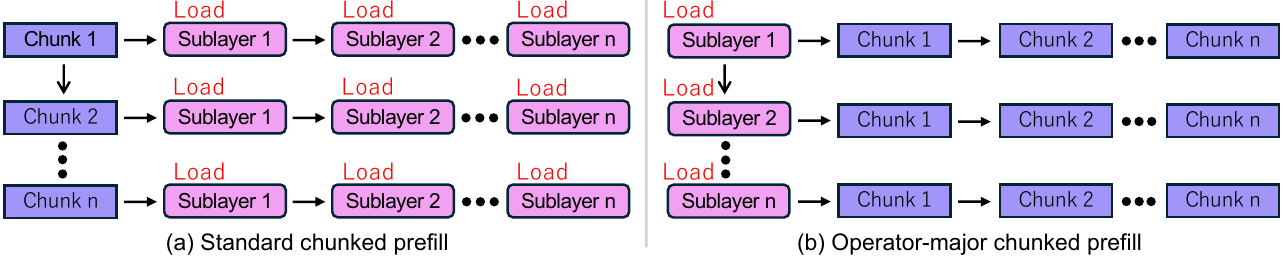}
  \vspace{-6mm}
  \caption{Comparison of standard and operator-major chunked prefill.}
  \label{fig:omcp}
  \vspace{-6mm}
\end{figure}

The Inference Runtime employs different execution strategies for prefill and decoding.
% During prefill, CPU computation is often slower than PCIe transfers, so the weights of each sublayer are transferred to the GPU as needed, and computation is performed on the GPU.
During prefill, Norn assigns all computation to the GPU, as CPU computation is often slower than PCIe transfers.
Latency is reduced by overlapping GPU computation with weight transfers for subsequent sublayers.
Norn optimizes weight transfers and activation memory usage through automatic transfer buffer sizing and operator-major chunked prefill (OMCP).

Norn determines the transfer buffer capacity required to overlap GPU computation with weight transfers based on the profiling results.
Let $C_A$ and $C_F$ denote the GPU computation times for attention and FFN, respectively, and $H_A$ and $H_F$ their weight transfer times.
With prefetching one sublayer ahead, the steady-state execution time per layer is $T_1=\max(C_A,H_F)+\max(C_F,H_A)$, while the theoretical lower bound under complete overlap between computation and transfers is $T_{\mathrm{lb}}=\max(C_A+C_F,H_A+H_F)$.
If $T_1=T_{\mathrm{lb}}$, prefetching one sublayer ahead is sufficient to achieve the lower bound.
Therefore, Norn prefetches one sublayer ahead when $T_1 \leq T_{\mathrm{lb}}$ and two sublayers ahead otherwise.
The transfer buffer capacity is calculated from the prefetch distance and weight sizes.
This ensures the minimum capacity required for prefetching while accounting for compute-transfer overlap efficiency.

For long inputs, the input sequence is divided into chunks to reduce activation memory usage.
As shown in Figure~\ref{fig:omcp}, standard chunked prefill~\cite{agrawal2024taming} processes all layers for each chunk, resulting in repeated transfers of offloaded weights.
To avoid these transfers, Norn employs OMCP, which processes all chunks for each sublayer before proceeding to the next sublayer.
% Layered Prefill~\cite{lee2026tokens} reduces repeated weight loading in chunked prefill through layer-wise processing, targeting MoE model serving.
Layered Prefill~\cite{lee2026tokens} reduces repeated loading of expert weights from off-chip GPU memory during chunked prefill through layer-group-wise processing in MoE model serving.
However, that paper reports lower performance than standard chunked prefill when applied to dense models.
In contrast, this paper demonstrates that sublayer-wise chunk processing can effectively reduce weight transfers even for dense models in a single-GPU offloading environment.

While standard chunked prefill retains only the hidden states of each chunk, OMCP must retain the hidden states of the entire input sequence to pass them to the next sublayer.
Let $N$ denote the input length, $M$ the chunk size, $d$ the hidden size, and $b$ the number of bytes per element.
The additional memory usage is approximately $(N-M)db$ bytes.
For example, with Qwen3-32B in FP16 at $N=128K$ and $M=1K$, activation memory usage increases by 1.24~GiB.
Thus, although OMCP requires more activation memory than standard chunked prefill, it avoids repeated weight transfers.
Nevertheless, it substantially reduces peak memory usage compared with prefill without chunking.

Chunk size affects prefill execution time and activation memory usage. 
A smaller chunk size reduces peak activation memory usage and makes it easier to retain the weights required by the initial decoding plan on the GPU after prefill, but may reduce compute-transfer overlap. %, exposing more weight-transfer overhead.
Therefore, Norn uses profiling results and estimated memory usage to evaluate the sum of prefill execution time and the weight transfer time required to transition to the initial decoding plan. 
It selects the chunk size with the lowest estimated execution time among candidates satisfying the GPU memory constraints. 
The KV cache generated during prefill is placed according to the device placements specified by the initial decoding plan, and the weights required by the initial plan are retained on the GPU to the extent permitted by GPU memory capacity. 
This reduces data transfers when transitioning from prefill to decoding.

During decoding, each sublayer is executed on the CPU or GPU specified by the execution plan. CPU execution uses data stored in host memory, while GPU execution uses either GPU-resident data or data transferred from host memory, depending on the placement of weights and the KV cache. Data not resident on the GPU is prefetched before it is needed by subsequent sublayers to overlap transfers with computation.
When the context length reaches a switching point, the execution plan is switched at a token boundary.
Based on the placement differences between the old and new plans, the runtime frees unnecessary GPU-resident weights, transfers newly required weights, relocates the KV cache, and adjusts the transfer buffer capacity, reusing existing GPU-resident data and memory regions wherever possible.

A shared CUDA stream is used for H2D transfers of weights and the KV cache, with transfers issued in sublayer execution order.
If the available buffer capacity is insufficient for an earlier transfer, subsequent transfers are not issued ahead of it, and execution waits until sufficient buffer space becomes available.
When the execution device changes between the CPU and GPU, activations are also transferred.
These transfers use a separate CUDA stream and are synchronized to satisfy data dependencies between sublayers.
In streaming execution, weights and the previously generated KV cache are retained in host memory, so only the newly generated KV cache is transferred from the GPU to the host memory.
This transfer is performed asynchronously on a dedicated CUDA stream after attention computation completes and is overlapped with subsequent computation.

\subsection{Implementation Details}

The Inference Runtime of Norn is implemented using PyTorch~\cite{NEURIPS2019_bdbca288} and Hugging Face Transformers~\cite{wolf2019huggingface}.
Norn is not tied to any specific dense model.
Norn is also a lossless method that does not involve quantization, although it can be combined with quantization.
For GPU attention computation, Norn uses PyTorch's scaled dot-product attention.
For CPU attention computation, Norn uses a custom kernel implemented in C++, ISPC, and OpenMP, based on the parallelization approach of NEO~\cite{jiang2025neo}.
This kernel parallelizes attention by partitioning the KV cache along the context dimension without replicating KV heads for GQA~\cite{ainslie2023gqa}.
Pinned memory is used for weights and the KV cache stored in host memory.
The weights of each sublayer are stored in a contiguous memory region and transferred together at the sublayer level.
The GPU-resident KV cache is not allocated upfront for the maximum context length.
Instead, its capacity is allocated based on the end of the plan's applicable range and the request's generation limit, and expanded as needed when switching plans.

\section{Evaluation}

\subsection{Experimental Setup}
\textbf{Testbeds.} 
We used two desktop systems for our experiments. 
Desktop~1 was equipped with an Intel Core Ultra 7 265K CPU, 128~GiB of main memory, and an NVIDIA RTX 5090 GPU with 32~GiB of memory. 
The CPU and GPU were connected via PCIe 5.0 $\times$16. 
Desktop 2 was equipped with an Intel Core i9-10980XE CPU, 128~GiB of main memory, and an NVIDIA RTX 3090 Ti GPU with 24~GiB of memory. 
The CPU and GPU were connected via PCIe 3.0 $\times$16. 
All experiments used 8 threads for CPU computation. 
\newline
\textbf{Models.}
We used Qwen3-14B and Qwen3-32B~\cite{yang2025qwen3} in FP16.
This allows evaluation under different memory requirements for weights and the KV cache.
% Using models of different sizes allows us to evaluate performance under varying memory requirements for model weights and the KV cache.
\newline
\textbf{Workloads.}
Following prior studies targeting interactive inference~\cite{chen2026fasttts,ukarande2026efficient,chen2025ktransformers}, we conducted our primary evaluation with a batch size of one. 
% We also evaluated performance with larger batch sizes in Appendix A.
For reproducibility and fair comparison, we used greedy decoding with a fixed output length.
Since our method preserves model accuracy and does not alter the operations performed, we evaluated performance solely as a function of input and output lengths.
We used dummy input sequences of the specified lengths.
The metrics were time to first token (TTFT), TPOT, and end-to-end latency (E2EL).
\newline
\textbf{Baselines.}
We compared Norn with Pipelined Sharding (PS)~\cite{ukarande2026efficient,ukarande2026pipeshardcode}, a state-of-the-art offloading method.
PS is built on llama.cpp and has two key parameters, \texttt{ctx-size} and \texttt{ubatch-size}.
The former specifies the maximum supported context length, while the latter specifies the maximum number of tokens processed in a single forward pass during prefill or decoding.
For a batch size of one, \texttt{ubatch-size} corresponds to the chunk size for chunked prefill.
To support diverse workloads, we set \texttt{ctx-size} to 128K, the maximum context length supported by Qwen3.
We also evaluated PS-O, an oracle variant that assumes the input and output lengths are known in advance and sets \texttt{ctx-size} for each request accordingly.
We set the chunk size to 1024, as commonly used in the Pipelined Sharding paper.

\subsection{Performance Across Input Lengths}

\begin{figure*}[t]
    \centering
    \begin{minipage}{0.49\textwidth}
        \includegraphics[width=\linewidth]{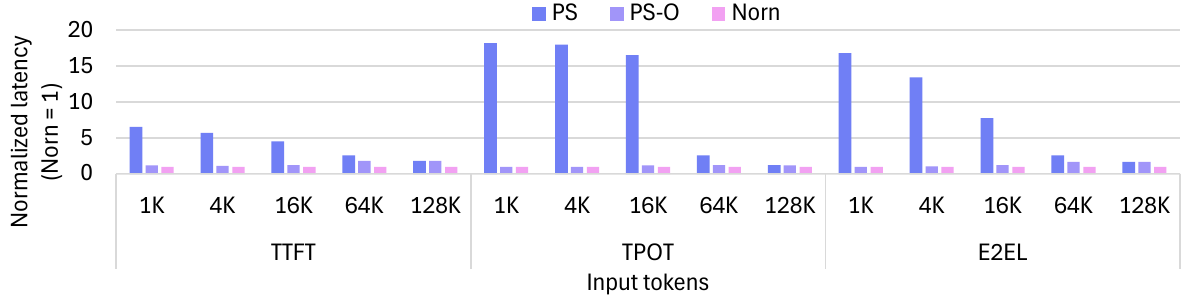}
        \vspace{-5mm}
        \subcaption{Qwen3-14B.}
        \label{fig:ex1-desktop1-14b}
    \end{minipage}
    \hfill % 画像間のスペース
    \begin{minipage}{0.49\textwidth}
        \includegraphics[width=\linewidth]{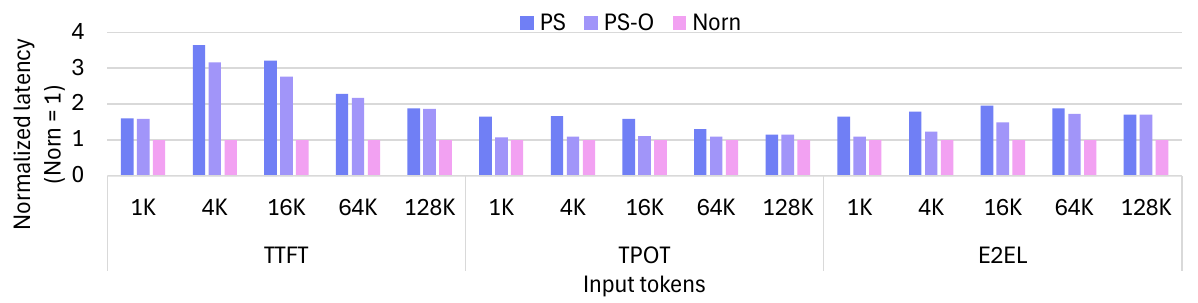}
        \vspace{-5mm}
        \subcaption{Qwen3-32B.}
        \label{fig:ex1-desktop1-32b}
    \end{minipage}
    \vspace{-3mm}
    \caption{Normalized TTFT, TPOT, and E2EL across input lengths on Desktop 1. The output length is fixed at 32 tokens.}%, and each metric is normalized to Norn.}
    \label{fig:ex1-desktop1}
    \vspace{-3mm}
\end{figure*}

\begin{figure*}[t]
    \centering
    \begin{minipage}{0.49\textwidth}
        \includegraphics[width=\linewidth]{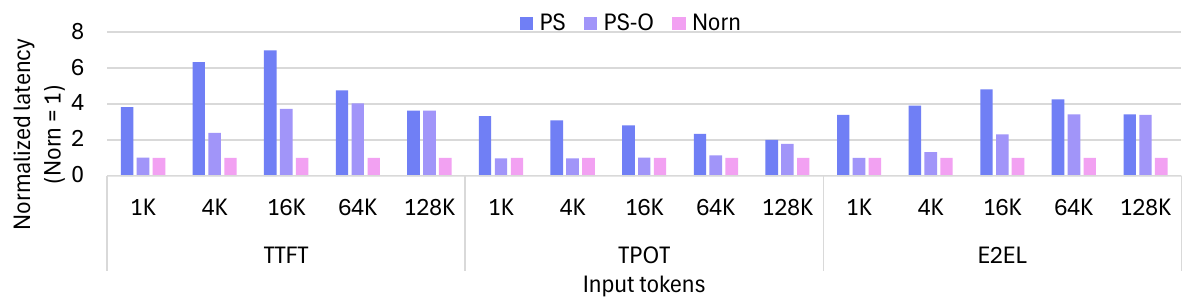}
        \vspace{-5mm}
        \subcaption{Qwen3-14B.}
        \label{fig:ex1-desktop2-14b}
    \end{minipage}
    \hfill % 画像間のスペース
    \begin{minipage}{0.49\textwidth}
        \includegraphics[width=\linewidth]{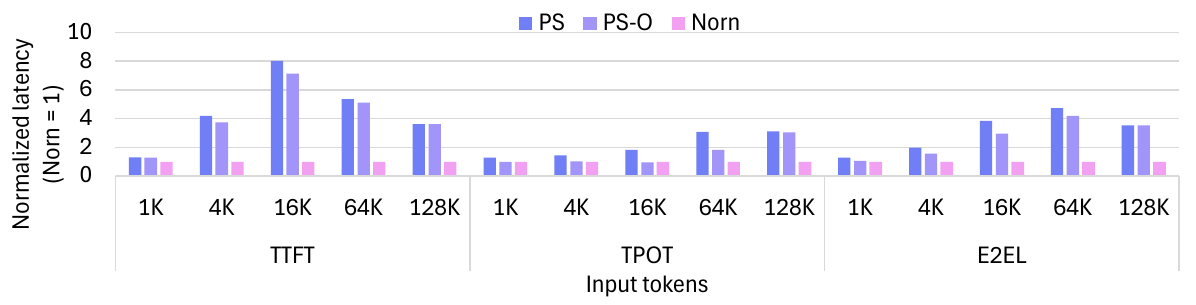}
        \vspace{-5mm}
        \subcaption{Qwen3-32B.}
        \label{fig:ex1-desktop2-32b}
    \end{minipage}
    \vspace{-3mm}
    \caption{Normalized TTFT, TPOT, and E2EL across input lengths on Desktop 2. The output length is fixed at 32 tokens.}%, and each metric is normalized to Norn.}
    \label{fig:ex1-desktop2}
    \vspace{-6mm}
\end{figure*}

Figures~\ref{fig:ex1-desktop1} and~\ref{fig:ex1-desktop2} show the results with the output length fixed at 32 tokens and the input length varied from 1K to 128K tokens.
Norn achieved lower E2EL than PS under all evaluated conditions, with up to a 16.8$\times$ speedup.
Compared with PS-O, which assumes that input and output lengths are known in advance, Norn achieved comparable performance for short inputs and up to a 4.2$\times$ speedup.
For TTFT, Norn outperformed both PS and PS-O under all evaluated conditions.
In particular, for Qwen3-32B on Desktop 2 with a 16K-token input, Norn achieved speedups of 8.0$\times$ and 7.1$\times$ over PS and PS-O, respectively.
PS and PS-O process each chunk through all layers in chunked prefill, repeatedly transferring offloaded weights for each chunk.
Although they prefetch the weights of the immediately following sublayer to the GPU to overlap computation with data transfers, the transfer overhead cannot always be fully hidden.
In contrast, Norn uses OMCP to process all chunks of each sublayer consecutively, avoiding repeated weight transfers.
Furthermore, Norn prefetches weights one or two sublayers ahead depending on computation and transfer times to increase the overlap between computation and data transfers.
These techniques reduce TTFT overhead from weight transfers.

For TPOT, PS, with its maximum supported context length fixed at 128K, exhibited substantial performance degradation, particularly for short inputs.
For Qwen3-14B on Desktop 1 with a 1K-token input, the TPOT of PS was 18.2$\times$ that of Norn.
In contrast, PS-O adjusts the allocated KV cache capacity according to input and output lengths, achieving performance close to that of Norn for short inputs.
However, PS-O also exhibited performance degradation for long inputs, with a TPOT 3.0$\times$ that of Norn at 128K tokens for Qwen3-32B on Desktop 2.
Two differences in placement and prefetching may contribute to Norn's performance advantage over PS-O.
First, PS-O determines placement based on a fixed rule that prioritizes GPU residency for attention weights and the KV cache, increasing the number of sublayers assigned to CPU execution at long context lengths.
With a 128K-token input, PS-O executed 32 FFN sublayers on the CPU, compared with 19 for Norn, for Qwen3-14B on Desktop 1.
Similarly, for Qwen3-32B on Desktop 2, PS-O executed 32 attention sublayers on the CPU, compared with 20 for Norn.
These results suggest that Norn reduces CPU computation by co-optimizing the placement of weights and the KV cache and the assignment of computation to the CPU or GPU.
Second, prefetching in PS-O is limited to the immediately following sublayer, whereas Norn can prefetch data for multiple subsequent sublayers.
This likely allows Norn to overlap CPU and GPU computation with data transfers more efficiently and make better use of streaming execution for weights and the KV cache.

\subsection{Decoding Performance Across Output Lengths}

% \begin{figure}[t]
%     \centering
%     \begin{minipage}{0.49\columnwidth}
%         \includegraphics[width=\linewidth]{fig/ex2_rtx5090_qwen14b.pdf}
%         \vspace{-5mm}
%         \subcaption{Qwen3-14B.}
%         \label{fig:ex2-desktop1-14b}
%     \end{minipage}
%     \hfill % 画像間のスペース
%     \begin{minipage}{0.49\columnwidth}
%         \includegraphics[width=\linewidth]{fig/ex2_rtx5090_qwen32b.pdf}
%         \vspace{-5mm}
%         \subcaption{Qwen3-32B.}
%         \label{fig:ex2-desktop1-32b}
%     \end{minipage}
%     \vspace{-3mm}
%     \caption{Normalized average TPOT across input and output lengths on Desktop 1.}
%     \label{fig:ex2-desktop1}
%     \vspace{-4mm}
% \end{figure}

% \begin{figure}[t]
%     \centering
%     \begin{minipage}{0.49\columnwidth}
%         \includegraphics[width=\linewidth]{fig/ex2_rtx3090ti_qwen14b.pdf}
%         \vspace{-5mm}
%         \subcaption{Qwen3-14B.}
%         \label{fig:ex2-desktop2-14b}
%     \end{minipage}
%     \hfill % 画像間のスペース
%     \begin{minipage}{0.49\columnwidth}
%         \includegraphics[width=\linewidth]{fig/ex2_rtx3090ti_qwen32b.pdf}
%         \vspace{-5mm}
%         \subcaption{Qwen3-32B.}
%         \label{fig:ex2-desktop2-32b}
%     \end{minipage}
%     \vspace{-3mm}
%     \caption{Normalized average TPOT across input and output lengths on Desktop 2.}
%     \label{fig:ex2-desktop2}
%     \vspace{-5mm}
% \end{figure}

\begin{figure}[t]
    \centering
    \begin{minipage}{0.49\columnwidth}
        \includegraphics[width=\linewidth]{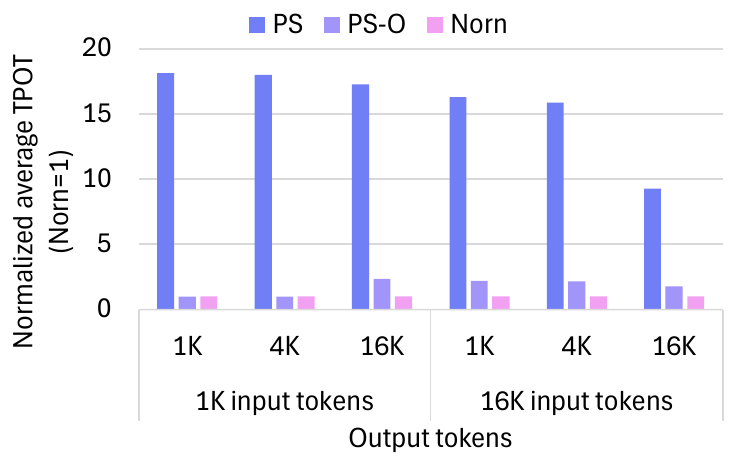}
        \vspace{-5mm}
        \subcaption{Qwen3-14B.}
        \label{fig:ex2-desktop1-14b}
    \end{minipage}
    \hfill
    \begin{minipage}{0.49\columnwidth}
        \includegraphics[width=\linewidth]{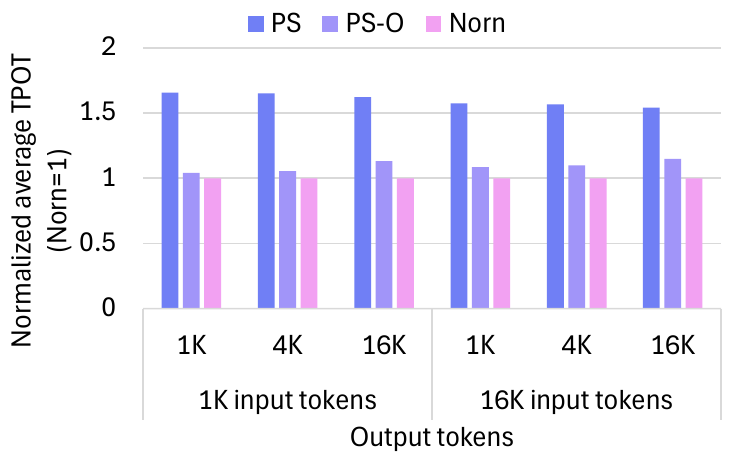}
        \vspace{-5mm}
        \subcaption{Qwen3-32B.}
        \label{fig:ex2-desktop1-32b}
    \end{minipage}
    \vspace{1mm} % 上段と下段の隙間
    \begin{minipage}{0.49\columnwidth}
        \includegraphics[width=\linewidth]{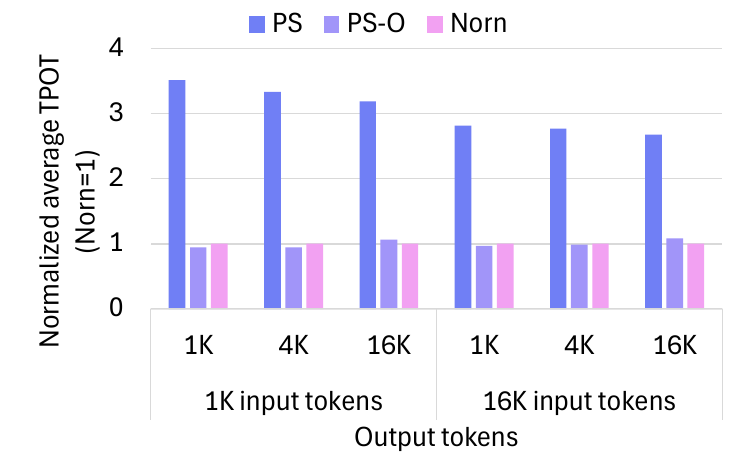}
        \vspace{-5mm}
        \subcaption{Qwen3-14B.}
        \label{fig:ex2-desktop2-14b}
    \end{minipage}
    \hfill
    \begin{minipage}{0.49\columnwidth}
        \includegraphics[width=\linewidth]{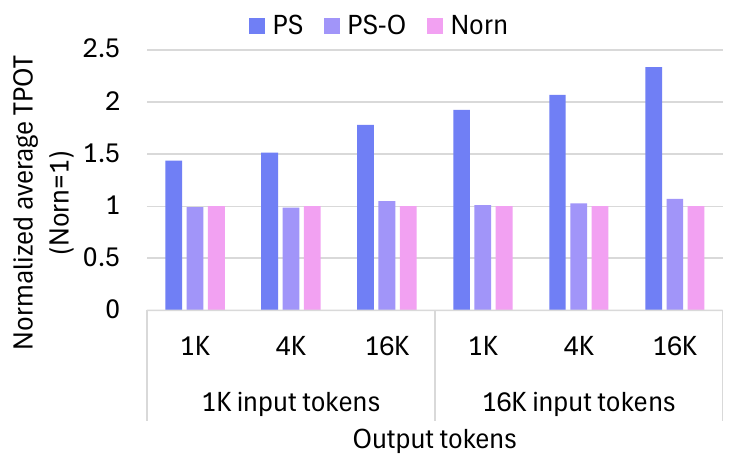}
        \vspace{-5mm}
        \subcaption{Qwen3-32B.}
        \label{fig:ex2-desktop2-32b}
    \end{minipage}
    \vspace{-3mm}
    \caption{Normalized average TPOT across input and output lengths on Desktop 1 (top) and Desktop 2 (bottom).}
    \label{fig:ex2}
    \vspace{-5mm}
\end{figure}

\begin{figure}[t]
  \centering
  \includegraphics[width=\columnwidth]{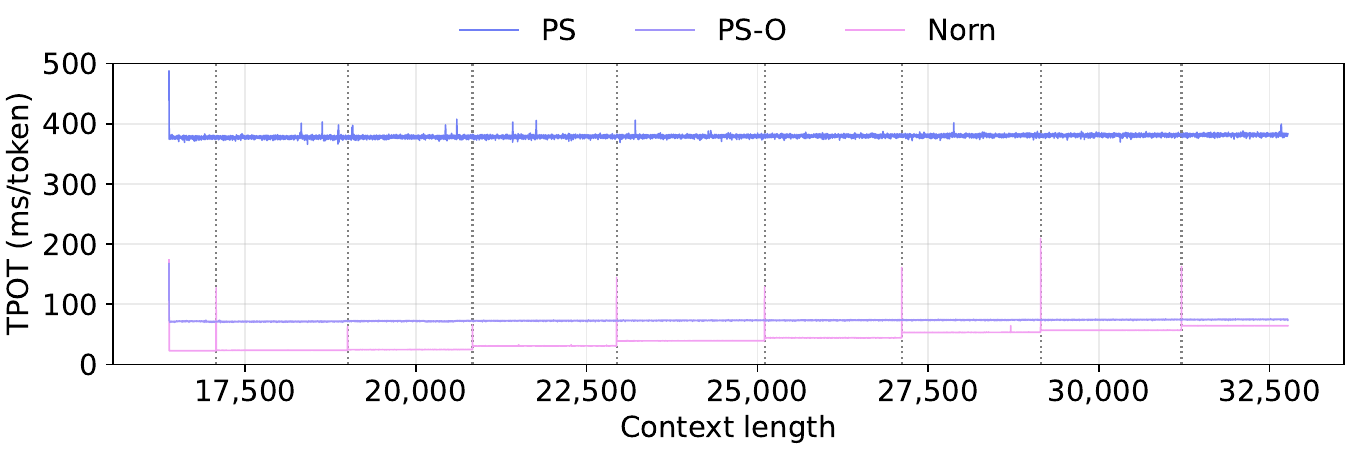}
  \vspace{-7mm}
  \caption{TPOT across context lengths for Qwen3-14B with 16K input and 16K output tokens on Desktop 1.}
  \label{fig:ex_tpot}
  \vspace{-4mm}
\end{figure}

Figure~\ref{fig:ex2} shows 
% Figures~\ref{fig:ex2-desktop1} and~\ref{fig:ex2-desktop2} show 
the average TPOT with input lengths fixed at 1K and 16K tokens and output lengths varied among 1K, 4K, and 16K tokens.
Norn outperformed PS under all evaluated conditions, achieving up to an 18.1$\times$ speedup.
Because PS uses a static placement configured for the maximum supported context length of 128K tokens, its performance degrades particularly at short context lengths.
Compared with PS-O, Norn achieved lower average TPOT in 16 of the 24 configurations, including all eight configurations with a 16K-token output, with up to a 2.33$\times$ speedup for Qwen3-14B on Desktop 1.
However, PS-O outperformed Norn in some configurations, particularly for short outputs, where Norn's average TPOT was up to 5.8\% higher.
Whereas PS-O sets the maximum supported context length based on known input and output lengths, Norn generates execution plans covering a wide range of context lengths without assuming the output length.
Consequently, Norn does not necessarily achieve the lowest average TPOT for every individual request.
Figure~\ref{fig:ex_tpot} shows TPOT as a function of context length for Qwen3-14B on Desktop 1 with a 16K-token input and a 16K-token output.
While PS and PS-O maintain nearly constant TPOT, Norn exhibits stepwise changes in TPOT due to execution plan switching as the context length increases.
Although temporary TPOT spikes occur at plan switches, Norn uses faster plans during the early stages of generation, achieving average TPOT speedups of 9.3$\times$ over PS and 1.8$\times$ over PS-O even after accounting for switching overhead.
These results demonstrate that context-length-aware dynamic placement effectively improves decoding performance for long outputs.

\subsection{Offline Planning Overhead}

\begin{table}[t]
\centering
\caption{Number of execution plans and planning time.}
\label{tab:planning_statistics}
\vspace{-3mm}
\begin{tabular}{lrr}
\toprule
Configuration & Final Plans & Total Time (s) \\
\midrule
Desktop 1 / Qwen3-14B & 27 & 1211 \\
Desktop 1 / Qwen3-32B & 2  & 245 \\
Desktop 2 / Qwen3-14B & 21 & 2081 \\
Desktop 2 / Qwen3-32B & 12 & 1643 \\
\bottomrule
\end{tabular}
\vspace{-4mm}
\end{table}

Table~\ref{tab:planning_statistics} shows the number of execution plans generated by the Dynamic Planner and the total planning time for a maximum context length of 128K tokens. 
The number of plans and the total planning time were 2--27 and 245--2,081~s, respectively.
The total planning time depends on the number of final plans, the number of times plans are generated for individual context-length ranges, and the time required for each generation. On Desktop 1, plans for individual ranges were generated 139 and 8 times for Qwen3-14B and Qwen3-32B, respectively, accounting for 64.1\% and 27.7\% of the total planning time. For Qwen3-14B, Desktop 2 required fewer plan generations (83), but the average generation time was longer than on Desktop 1 (13.1~s versus 5.6~s), resulting in a longer total planning time.
We used HiGHS~\cite{huangfu2018parallelizing} to solve the MILP problems. On Desktop 1, 3,248 and 1,152 MILP instances were solved for Qwen3-14B and Qwen3-32B, respectively, all to optimality. The average solve times were 84.5~ms and 33.5~ms, accounting for 22.7\% and 15.7\% of the total planning time, respectively.

\subsection{Effectiveness of Operator-Major Chunked Prefill}

\begin{figure}[t]
    \centering
    \begin{minipage}{0.47\columnwidth}
        \includegraphics[width=\linewidth]{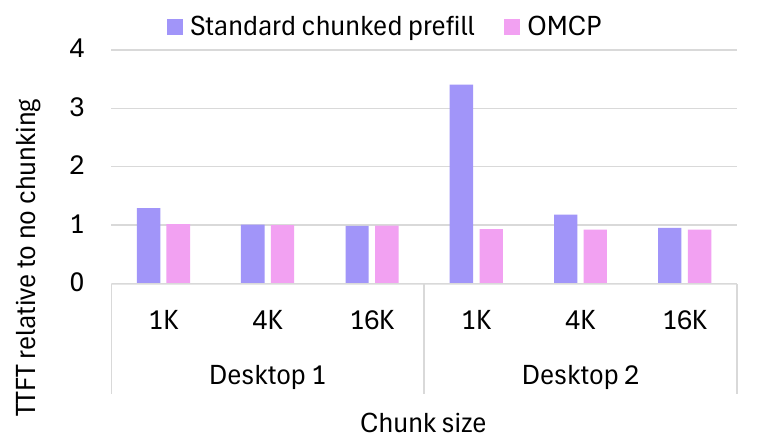}
        \vspace{-6mm}
        \caption{Normalized TTFT for Qwen3-14B with a 64K-token input.}
        \label{fig:ex_omcp_ttft}
    \end{minipage}
    \hfill % 画像間のスペース
    \begin{minipage}{0.47\columnwidth}
        \includegraphics[width=\linewidth]{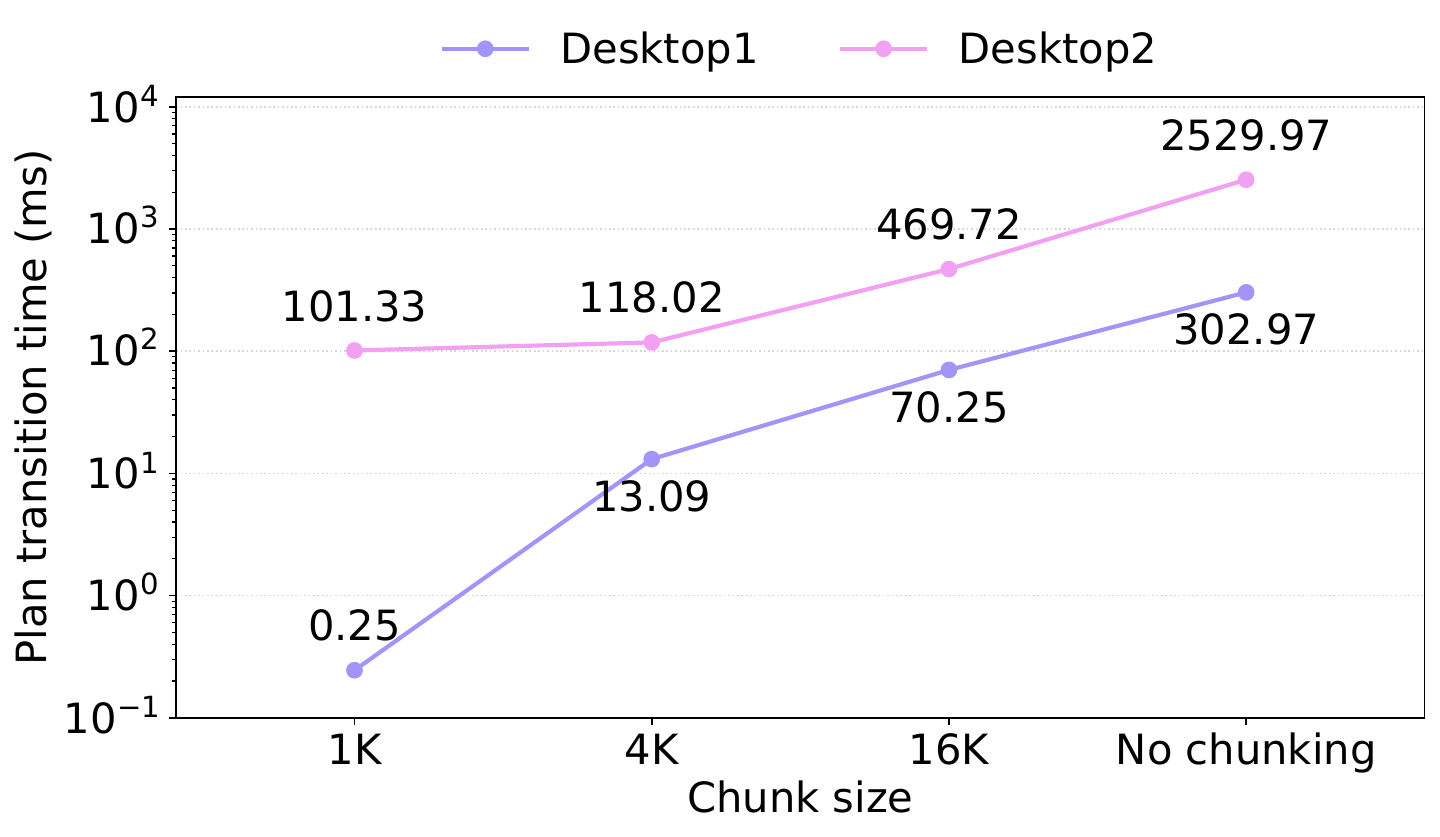}
        \vspace{-6mm}
        \caption{Plan transition time for Qwen3-14B with a 64K-token input.}
        \label{fig:ex_plan_trans}
    \end{minipage}
    \vspace{-5mm}
\end{figure}

We evaluated Qwen3-14B on Desktops 1 and 2 with a fixed input length of 64K tokens, varying the chunk size among 1K, 4K, and 16K, and measured TTFT and plan transition time.
Figure~\ref{fig:ex_omcp_ttft} shows TTFT normalized to the no-chunking configuration.
With a 1K chunk size on Desktop 2, standard chunked prefill increased TTFT to 3.41$\times$ the baseline, whereas OMCP achieved 0.93$\times$, providing a 3.66$\times$ speedup over the standard method.
On Desktop 1, OMCP achieved 1.02$\times$, compared with 1.29$\times$ for the standard method.
This is because OMCP reuses weights across chunks and avoids repeated transfers, making it particularly effective on Desktop 2, where PCIe bandwidth is limited.
As the chunk size increases, the performance gap between the two methods narrows because longer computation times allow standard chunked prefill to hide weight transfer overhead.
Figure~\ref{fig:ex_plan_trans} shows the transition time from prefill to the initial decoding plan for Qwen3-14B.
With a 1K chunk size, OMCP reduced the transition time from 303~ms without chunking to 0.25~ms on Desktop 1 and from 2,530~ms to 101~ms on Desktop 2.
The transition time decreased as the chunk size became smaller.
This is because smaller chunks reduce peak activation memory usage, allowing more weights and the KV cache required by the initial decoding plan to remain in GPU memory.
These results demonstrate that OMCP reduces the transition time to initial decoding while limiting TTFT overhead even with small chunk sizes.

\section{Related Work}

Various methods have been proposed to accelerate LLM inference in resource-constrained environments by jointly utilizing the CPU and GPU.
HeteGen~\cite{zhao2024hetegen} and TwinPilots~\cite{yu2024twinpilots} reduce inference latency by overlapping CPU computation, GPU computation, and data transfers.
FlexInfer~\cite{na2025flexinfer} and Pipelined Sharding~\cite{ukarande2026efficient} determine computation assignments and placement based on sequence length and hardware characteristics.
Layer-wise Scheduling~\cite{sugiura2026layer} assigns each layer to CPU execution, GPU resident execution, or GPU streaming execution and overlaps weight transfers with CPU computation.
However, because KV cache placement depends on the execution device of each layer, weights and the KV cache cannot be placed independently.
% These methods do not focus on adapting the placement of weights and the KV cache within a single request as the KV cache grows during decoding.
These methods do not focus on adapting weight and KV cache placement as the KV cache grows during decoding.

PowerInfer~\cite{song2024powerinfer,xue2024powerinfer2fastlargelanguage} and LLM in a Flash~\cite{alizadeh2024llm} exploit activation sparsity to reduce computation and data transfers by predicting which neurons will be activated.
However, these methods require predictors to identify active neurons, limiting their effectiveness for models with insufficient activation sparsity.
Furthermore, prediction errors may degrade inference quality.
FlexGen~\cite{sheng2023flexgen} and TightLLM~\cite{hu2025tightllm} target offline batch inference and maximize throughput by amortizing weight transfers across multiple requests at the expense of latency.
NEO~\cite{jiang2025neo} and APEX~\cite{fan2026apex} target online serving and improve throughput by offloading the KV cache and part of attention computation to the CPU and executing CPU and GPU computations in parallel, but assume multiple concurrent requests.
HeadInfer~\cite{luo2025headinfer} and InfiniGen~\cite{lee2024infinigen} propose methods to reduce KV cache memory usage and transfer overhead but do not jointly optimize GPU memory allocation between weights and the KV cache.

\section{Conclusion}

In this paper, we proposed Norn, an offloaded LLM inference method that dynamically switches execution plans according to context length. 
Norn precomputes multiple execution plans with optimized assignments of computation to the CPU or GPU and placements of weights and the KV cache, along with their switching points, to adapt to increasing context length during decoding. 
It also employs operator-major chunked prefill to avoid repeated weight transfers while reducing memory usage. 
Experiments with Qwen3-14B and Qwen3-32B on two desktop systems showed that Norn achieved up to a 16.8$\times$ speedup over Pipelined Sharding and up to a 4.2$\times$ speedup over its oracle configuration in which input and output lengths are known in advance.

% \section*{Acknowledgements}

% \textbf{Do not} include acknowledgements in the initial version of
% the paper submitted for blind review.


\bibliography{example_paper}
\bibliographystyle{mlsys2025}

\appendix
% \input{7_Appendix}
% \section{Please add supplemental material as appendix here}
% %
% Put anything that you might normally include after the references as an appendix here, {\it not in a separate supplementary file}. Upload your final camera-ready as a single pdf, including all appendices.

%%%%%%%%%%%%%%%%%%%%%%%%%%%%%%%%%%%%%%%%%%%%%%%%%%%%%%%%%%%%%%%%%%%%%%%%%%%%%%%
%%%%%%%%%%%%%%%%%%%%%%%%%%%%%%%%%%%%%%%%%%%%%%%%%%%%%%%%%%%%%%%%%%%%%%%%%%%%%%%

\end{document}